\documentclass[aps,twocolumn,preprintnumbers,final]{revtex4}

\usepackage{graphicx,amsmath,amssymb,mathrsfs,array,longtable,delarray,verbatim,epsfig}
\graphicspath{{E:\My_papers\Compressibility_PRL}}

\begin{document}

\title{Compressibility measurements of quasi-one-dimensional quantum wires}

\author{L. W. Smith$^{1,*}$, A. R. Hamilton$^2$, K. J. Thomas$^{3,\dag}$, M. Pepper$^4$, I. Farrer$^1$, J. P. Griffiths$^1$, G. A. C. Jones$^1$, D. A. Ritchie$^1$}

\affiliation{
$^{1}$Cavendish Laboratory, J. J. Thomson Avenue, Cambridge, CB3 OHE, United Kingdom\\
$^{2}$School of Physics, The University of New South Wales, Sydney 2052, Australia\\
$^{3}$Department of Electronic and Electrical Engineering, Sungkyunkwan University, Suwon 440-746, South Korea\\
$^{4}$Department of Electronic and Electrical Engineering, University College London, London, WC1E 7JE, United Kingdom}

\begin{abstract}

We report measurements of the compressibility of a one-dimensional (1D) quantum wire, defined in the upper well of a GaAs/AlGaAs double quantum well heterostructure. A wire defined simultaneously in the lower well probes the ability of the upper wire to screen the electric field from a biased surface gate. The technique is sensitive enough to resolve spin-splitting of the subbands in the presence of an in-plane magnetic field. We measure a compressibility signal due to the 0.7 structure and study its evolution with increasing temperature and magnetic field. We see no evidence of the formation of the quasibound state predicted by the Kondo model, instead our data are consistent with theories which predict that the 0.7 structure arises as a result of spontaneous spin polarization.

\end{abstract}

\maketitle

The conductance of a quasi-one-dimensional (1D) quantum wire~\cite{thornton86}, quantized in units of $2e^2/h$~\cite{wharam88, vanWees88}, is in general well described by a non-interacting single-particle picture. However, below the first conductance plateau, a feature close to 0.7($2e^2/h$) exists~\cite{thomas96}, believed to arise from electron interactions. Many theories have been proposed as to why this `0.7 structure' occurs, including spontaneous spin polarization~\cite{thomas96, wang96}, the Kondo effect~\cite{cronenwett02, meir02}, and Wigner crystallization~\cite{klironomos06, matveev04}. Its cause has yet to be conclusively determined, and this topic remains one of significant interest~\cite{thomas96, wang96, cronenwett02, meir02, klironomos06, matveev04, reilly02, bird04}.

To date, most experimental studies of 1D systems have focused on measuring conductance through the quantum wire. Notable exceptions include thermopower~\cite{appleyardPRB2000, molenkampPRL1990} and shot-noise~\cite{reznikovPRL95, rochePRL04} measurements, but, like conductance, these are thermodynamically non-equilibrium measurements. We measure the compressibility of a 1D wire, which, as a thermodynamic property, measures the equilibrium properties of the system.

The compressibility ($\kappa$) of a two-dimensional electron gas (2DEG) is given by $\kappa^{-1}=N^2\partial\mu/\partial N$, where $\mu$ is the chemical potential and $N$ is the areal density. The compressibility is directly related to the screening ability of the electron gas, such that a highly compressible system more easily screens local electric fields. This can be measured using two closely-separated electron layers in a double quantum well (DQW) structure, first shown by Eistenstein \emph{et al.}~\cite{eisenstein92, eisenstein94}. Compressibility measurements are a useful tool for measuring interaction effects, for example in the fractional quantum Hall regime~\cite{eisenstein92}, and have recently been used to study electronic states in graphene~\cite{martinNature08}.

In early compressibility studies of 1D systems, arrays of quantum wires were investigated using capacitive techniques~\cite{Drexler1994,Smith1987}. Later, Castleton~\emph{et al.}~\cite{castleton} measured an individual quantum wire, which showed enhanced compressibility as each 1D subband was populated. More recently, L\"{u}scher~\emph{et al.}~\cite{luscher} predicted a compressibility feature associated with the formation of a quasibound state in the 1D channel,  using density functional theory (DFT) calculations.

In this Letter we present measurements of the compressibility of a 1D wire in zero and finite in-plane magnetic field. We resolve the spin splitting of 1D subbands in the compressibility signal, due to the high sensitivity of our technique. As the magnetic field is increased, the compressibility signal shows a peak associated with the lowest spin-split subband. A very similar peak related to the 0.7 structure emerges as a function of temperature at zero field. Our results support earlier studies which suggest that the origin of the 0.7 structure is related to spontaneous spin polarization~\cite{thomas96}. We do not observe the features predicted in Ref.~\cite{luscher} for a Kondo-like origin of the 0.7 structure, despite the sensitivity of our measurement.

\begin{figure}[th]
\begin{center}
\includegraphics[scale=0.33]{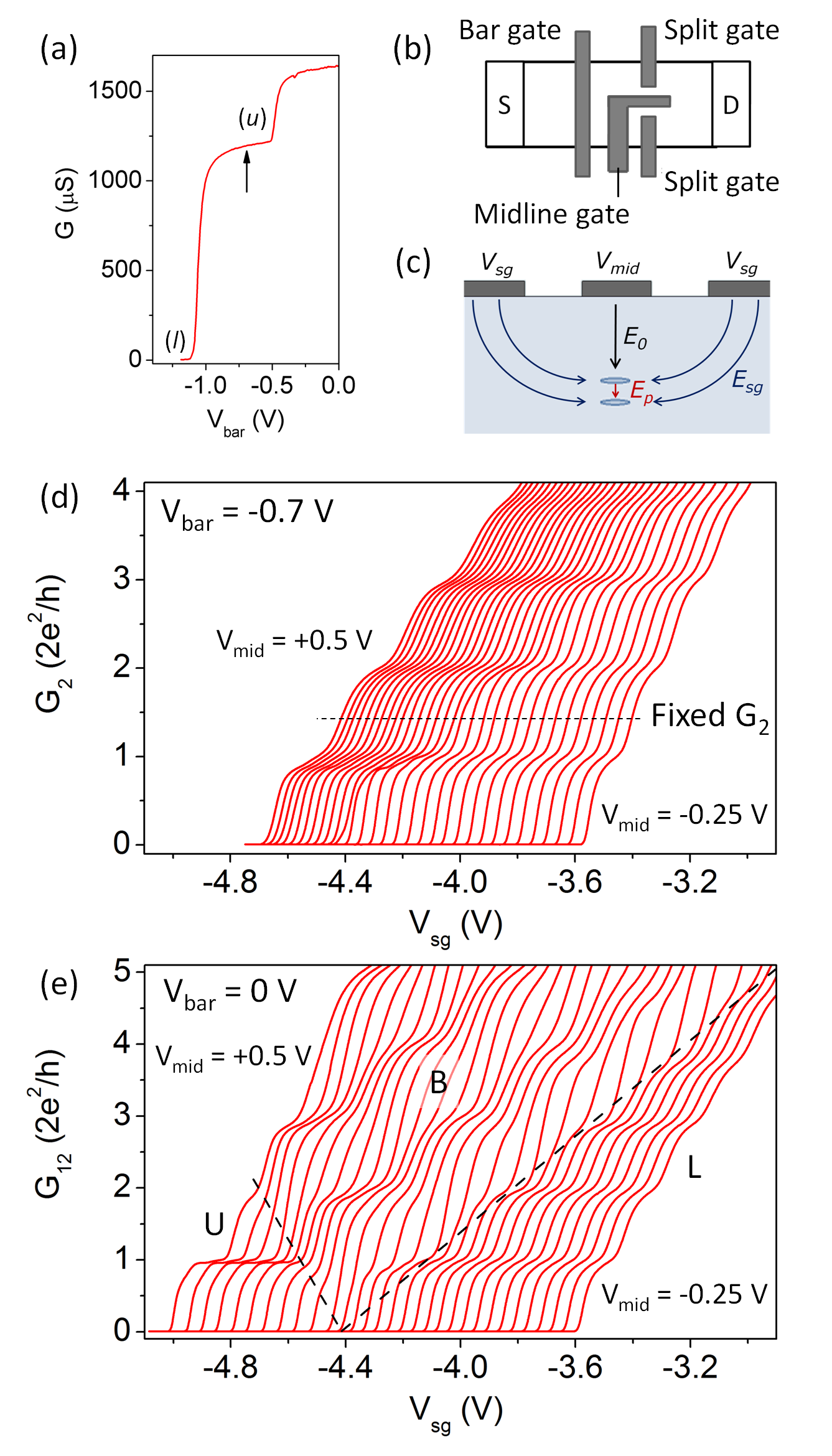}
\end{center}
\caption{(a): Conductance as a function of $V_{bar}$. Labels (\emph{u}) and (\emph{l}) mark the depletion of the upper and lower electron layers, respectively. The arrow indicates $V_{bar}=-0.7$ V, typically applied to fully deplete the upper layer. (b): Device schematic, showing the bar gate, split gates and midline gate. The source and drain ohmic contacts are marked by S and D, respectively. (c): Simplified diagram of a cross section through the device, showing the electric fields generated by biasing the split gates ($E_{sg}$), the midline gate ($E_0$), and the field which penetrates the upper wire ($E_p$). (d): Conductance measured through the lower wire alone ($G_2$) as a function of $V_{sg}$ ($V_{bar}=-0.7$ V). The horizontal dashed line indicates a fixed $G_2$. (e): Conductance through both wires ($G_{12}$) as a function of $V_{sg}$, ($V_{bar}=0$ V). In both (d) and (e) $V_{mid}$ is stepped from 0.5 V on the left to -0.25 V on the right, in 25 mV intervals. Data are from a typical device, measured at 300 mK, and corrected for series resistance.}
\label{fig1}
\end{figure}

Our samples were fabricated on DQW GaAs/AlGaAs heterostructures, in which two 15 nm-wide quantum wells are separated by a 30 nm AlGaAs barrier. The center of the quantum wells are 292.5 and 337.5 nm below the surface of the wafer. Split-gate devices with a midline gate were patterned using electron-beam lithography, as shown schematically in Fig.~\ref{fig1}(b). The split gates were 0.5 $\mu$m long and 0.8 $\mu$m wide, and the midline gate was 0.3 $\mu$m wide, positioned at the center of split gate. The combination of split and midline gates allow 1D channels to be defined in both layers simultaneously~\cite{castleton}. Applying a voltage on the bar gate [see Fig.~\ref{fig1}(b)] depletes electrons from the top layer, allowing the lower layer to be independently contacted. Fig.~\ref{fig1}(a) shows the successive depletion of the upper and lower 2D layers as a function of bar-gate voltage ($V_{bar}$), marked by \emph{(u)} and \emph{(l)}, respectively. When $V_{bar}=0$ we measure the conductance of both layers ($G_{12}$), and for $V_{bar}=-0.7$ V (marked by the arrow), we measure the conductance of the lower layer alone ($G_2$).

\begin{figure}[t]
\begin{center}
\includegraphics[scale=0.3]{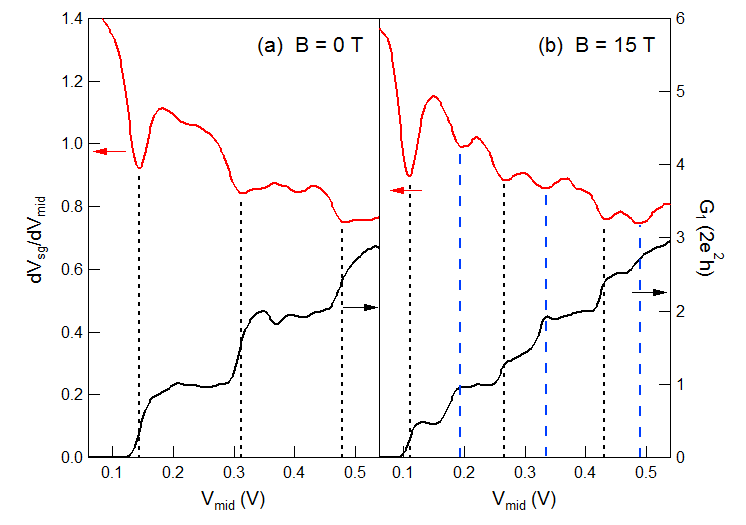}
\end{center}
\caption{(a): Compressibility signal $\mathrm{d}V_{sg}/\mathrm{d}V_{mid}$ against $V_{mid}$ (red trace, left axis), for $G_2\approx1.5(2e^2/h)$ ($B=0$ T). The conductance of the upper wire ($G_1$) is plotted in black (right axis). Minima in $\mathrm{d}V_{sg}/\mathrm{d}V_{mid}$ coincide with inter-plateau rising edges, as illustrated by the dashed lines. The equivalent data at $B = 15$ T is shown in (b), for $G_2\approx1.75(2e^2/h)$. The lifting of the spin degeneracy results in the appearance of spin-split plateaux in $G_1$, and we resolve additional minima in $\mathrm{d}V_{sg}/\mathrm{d}V_{mid}$ associated with the spin-splitting of subbands (marked by the long-dashed lines). This measurement was performed at 25 mK.}
\label{fig2}
\end{figure}

\begin{figure}[t]
\begin{center}
\includegraphics[scale=0.27]{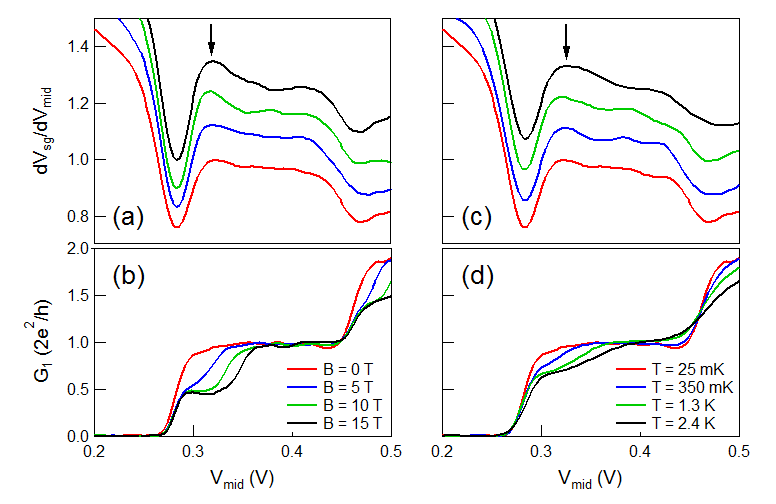}
\end{center}
\caption{(a) and (b): Magnetic field dependence of $\mathrm{d}V_{sg}/\mathrm{d}V_{mid}$ and corresponding $G_1$, respectively, for $B=0, 5, 10$ and 15 T. $B=15$ T data are shifted horizontally by 6.5 mV to align with other data. Measurements were performed at $T=25$ mK. In (b) data are vertically offset for clarity; the arrow marks the peak which develops as $B$ increases. (c) and (d): Temperature dependence of $\mathrm{d}V_{sg}/\mathrm{d}V_{mid}$ and corresponding $G_1$, respectively, at $T=25$ mK, $350$ mK, $1.3$ K and 2.4 K. $T= 350$ mK, $1.3$ K and $2.4$ K data are shifted horizontally by 2.5, 5, and -3.5 mV, respectively, to align with $25$ mK data~\cite{corr}. In (c) $\mathrm{d}V_{sg}/\mathrm{d}V_{mid}$ are vertically offset for clarity; the peak which develops with increasing temperature is marked by the arrow. Temperature-dependent measurements were performed at $B=0$ T. $G_2 = 0.23(2e^2/h)$ for both data sets~\cite{G2}.}
\label{fig3}
\end{figure}

\begin{figure}[t]
\begin{center}
\includegraphics[scale=0.3]{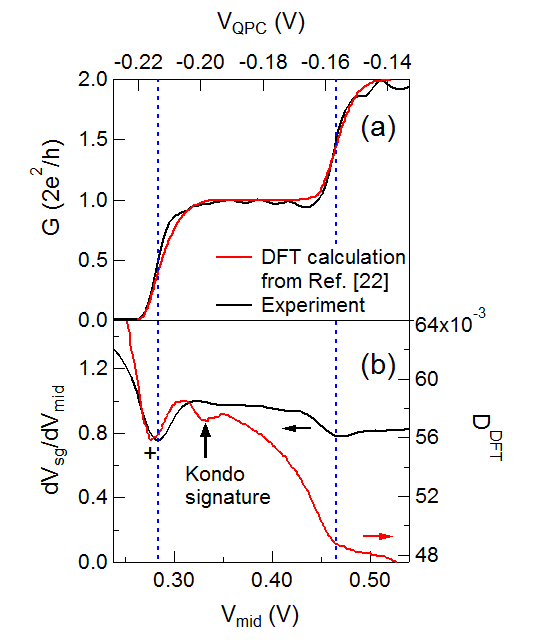}
\end{center}
\caption{(a): Calculated conductance of the quantum point contact (QPC) from L\"{u}scher \emph{et al.}~\cite{luscher} (red traces, top axis), compared with $G_1$ at $B=0$ T and $T=25$ mK (black trace, bottom axis). Horizontal axes are scaled to align the conductance data. (b): Corresponding derivatives $D^{DFT}$ (from Ref.~\cite{luscher}, red trace, right and top axes), and $\mathrm{d}V_{sg}/\mathrm{d}V_{mid}$ (experimental data, black trace, left and bottom axes). The two $y$ axes are equivalent~\cite{noteDFT}, and are scaled such that the depth of the first minima, marked by +, is of the same magnitude. The secondary dip in $D^{DFT}$, marked by the vertical arrow, was predicted to be a signature of the formation of a quasibound state. The vertical dashed lines mark transitions between the plateaux as 1D subbands are populated.}
\label{fig4}
\end{figure}

Crucial to this experiment is the quality of devices and stability of the 2DEG. Over 20 devices were tested at 300 mK. Devices which exhibited clean, stable and reproducible conductance characteristics were selected for further measurement. The data presented in Figs.~\ref{fig2}-\ref{fig4} are from a device for which the total carrier density and mobility were measured to be $2.4\times10^{11}$ cm$^{-2}$ and $3\times10^6$ cm$^2$V$^{-1}$s$^{-1}$, respectively (all measurements were performed in the absence of illumination). We measured similar results in two further devices.  Data presented were measured using a dilution refrigerator with a base temperature of 25 mK, and a $^3$He system at 300 mK. A magnetic field was applied in the plane of the 2DEG, perpendicular to the transport direction.

Fig.~\ref{fig1}(d) shows the conductance of the lower wire ($G_2$) as a function of $V_{sg}$. The bar gate is biased so that current only flows through the lower wire, even if the upper wire is occupied. For each trace $V_{mid}$ is fixed, and between traces $V_{mid}$ is stepped by $25$ mV, from $0.5$ V on the left to -0.25 V on the right. When the bar-gate bias is removed, the total conductance ($G_{12}$) is measured, shown in Fig.~\ref{fig1}(e). Regions are identified in which either the upper wire alone is populated (U), both wires are populated simultaneously (B), or the lower wire alone is populated (L)~\cite{castleton}. The dashed lines mark a `V' separating these regions. On the left and right of the V conductance is quantized in $2e^2/h$ multiples, indicating single-wire transport. Conductance in the middle region is quantized in $4e^2/h$ multiples, due to transport through the two wires in parallel. The upper-wire conductance ($G_1$) can be extracted from these data, since $G_1 = G_{12} - G_2$. We do not observe coupling between the two wires, as seen in devices with more closely-spaced quantum wells~\cite{thomas99}.

To determine the compressibility of the upper 1D wire, we measure how well it screens the electric field from the midline gate. This is accomplished by changing the electric field incident upon the upper wire ($\delta E_0$) from the midline gate, and measuring the corresponding change in the electric field ($\delta E_p$) which reaches the lower wire. The electric fields generated by biasing the surface gates are shown in the schematic cross-section of the device in Fig.~\ref{fig1}(c). The $\delta E_p$ causes a change in the carrier density, and hence conductance $G_2$, of the lower wire. This can be compensated for by adjusting $V_{sg}$ by $\delta V_{sg}$, so that the total electric field incident on the lower wire remains fixed (thereby keeping $G_2$ at a constant value~\cite{fixedG2}). Since $\delta E_p$ depends on the screening ability of the upper wire, the compressibility is directly related to the measured $\mathrm{d}V_{sg}/\mathrm{d}V_{mid}$. For example, if the top wire is highly compressible, $V_{mid}$ is well screened, and little field penetrates onto the lower wire: $G_2$ changes by a small amount in response to $V_{mid}$, and $\mathrm{d}V_{sg}/\mathrm{d}V_{mid}$ is also small. Therefore, minima in $\mathrm{d}V_{sg}/\mathrm{d}V_{mid}$ correspond to enhanced compressibility. We ignore the direct capacitive coupling between the midline gate and the lower 1D channel, which adds a rising background to the measured signal as $V_{mid}$ decreases.

Fig.~\ref{fig2}(a) shows the compressibility signal $\mathrm{d}V_{sg}/\mathrm{d}V_{mid}$ (left axis) and the corresponding conductance of the upper layer $G_1$ (right axis), at $B=0$ T. The $\mathrm{d}V_{sg}/\mathrm{d}V_{mid}$ trace is extracted from data similar to Fig.~\ref{fig1}(d), where $\delta V_{sg}$ is the spacing between neighbouring traces at fixed $G_2$ [shown by the horizontal dashed line in Fig.~\ref{fig1}(d)], and $\delta V_{mid}$ is the amount that $V_{mid}$ is stepped between traces. In Fig.~\ref{fig2}(a) we observe minima in $\mathrm{d}V_{sg}/\mathrm{d}V_{mid}$ which coincide with the population of 1D subbands in the upper wire. This is highlighted by the vertical dashed lines, which align with transitions (risers) between plateaux in $G_1$. At these minima $V_{mid}$ is screened most effectively by the upper wire, and therefore the compressibility is maximum. This occurs when the density of states is large (i.e. at the subband edges). This measurement was performed at 25 mK, for $G_2\approx1.5(2e^2/h)$~\cite{sensitiv}.

Fig.~\ref{fig2}(b) shows the same measurement at $B = 15$ T. Additional plateaux in $G_1$ confirm that the 1D subbands are spin split, due to the Zeeman energy. Secondary minima appear in $\mathrm{d}V_{sg}/\mathrm{d}V_{mid}$, which align with the beginning of spin-split plateaux in $G_1$ (indicated by the long-dashed lines). The ability to resolve spin splitting in the compressibility signal highlights the sensitivity of our technique, since the Zeeman energy is small ($\approx10\%$ of the 1D subband spacing). It is interesting to note that at $B = 0$ T the minima in $\mathrm{d}V_{sg}/\mathrm{d}V_{mid}$ align with the center of risers in $G_1$, whereas at $B = 15$ T, the additional minima in $\mathrm{d}V_{sg}/\mathrm{d}V_{mid}$ due to the spin splitting occur at higher energies (larger $V_{mid}$, tending to align with the beginning of plateaux).

We focus on the magnetic-field dependent behavior of the lowest subband. Fig.~\ref{fig3}(a) shows the compressibility signal $\mathrm{d}V_{sg}/\mathrm{d}V_{mid}$ at $B=0, 5, 10$ and $15$ T (vertically offset for clarity); corresponding $G_1$ traces are given in Fig.~\ref{fig3}(b). Significantly, a peak emerges in the compressibility signal as $B$ increases, marked by the arrow in Fig.~\ref{fig3}(a). In other experiments, a small peak was observed as low as $B=2$ T, which was shown to evolve smoothly into a large peak at higher fields, while a clear 0.7 structure in $G_1$ developed into the 0.5 plateau. 

The peak is clearly related to the onset of spin polarization, as it develops in line with the $0.5(2e^2/h)$ plateau in $G_1$. It represents a reduction in the density of states, which is consistent with the fact that at high fields, only the lowest spin subband is populated initially (instead of two spin-degenerate subbands in the zero-field case). 

We now examine the temperature dependence of the compressibility. Fig.~\ref{fig3}(c) shows the compressibility signal $\mathrm{d}V_{sg}/\mathrm{d}V_{mid}$ at $T=25$ mK, $350$ mK, $1.3$ K and $2.4$ K, for $B=0$ T (vertically offset for clarity). Corresponding $G_1$ traces are given in Fig.~\ref{fig3}(d), and data are horizontally aligned by no more than $5$ mV~\cite{corr}. In Fig.~\ref{fig3}(c), a peak in the compressibility signal emerges as temperature increases (marked by the arrow). At the same time a 0.7 structure develops in $G_1$~\cite{thomas96}. The compressibility response as temperature increases is strikingly similar to the compressibility response as magnetic field is increased at $T=25$ mK -- data in Figs.~\ref{fig3}(a) and (c) are almost identical.

We note that while the compressibility data in Figs.~\ref{fig3}(a) and (c) are so similar, the conductance data in Figs.~\ref{fig3}(b) and (d) are markedly different. This gives us confidence that we are performing a true thermodynamic measurement; $\mathrm{d}V_{sg}/\mathrm{d}V_{mid}$ is not simply reflecting changes in $G_1$, but is rather measuring the density of states in the wire.

We now consider how the compressibility data fits with a Kondo-like scenario, by comparing our results with density functional theory (DFT) calculations~\cite{noteDFT} performed by L\"{u}scher \emph{et al.}~\cite{luscher}. Figs.~\ref{fig4}(a) and (b) show our data at $B=0$ T and $T=25$ mK (black traces) with the simulations from Ref.~\cite{luscher} (red traces). We aligned the horizontal axes of our measured data with the simulations from Ref.~\cite{luscher} so that the conductance traces overlapped, as shown in Fig.~\ref{fig4}(a). Vertical axes in Fig.~\ref{fig4}(b) were aligned such that the depth of the first minima (marked by +), was equal for the experimental and simulated data, to allow an approximate comparison of our data with the only theoretical calculations currently available.

The DFT calculations predict that the formation of the quasibound state necessary for the Kondo effect is signalled by a characteristic dip in the compressibility signal, marked by the vertical arrow in Fig.~\ref{fig4}(b). Our experimental data is more sensitive than the simulated response, indicated by better defined minima in our compressibility signal when a subband in the upper wire is populated. Despite this, we do not observe the secondary dip that was predicted to be the signature of a quasibound state. Therefore, our results do not appear to support a Kondo-like scenario~\cite{highT}.

In contrast to the Kondo model, it has been proposed that the 0.7 structure occurs due to spontaneous spin polarization. This conclusion is primarily motivated by the continuous evolution of the 0.7 structure into the $e^2/h$ plateau as magnetic field increases~\cite{thomas96}. In the current experiment, we measure the compressibility response of the 0.7 structure by increasing the temperature, and find that its compressibility response is very similar to that as magnetic field is increased. Indeed, the integrated area under the peak in $\mathrm{d}V_{sg}/\mathrm{d}V_{mid}$ changes by less than $3$ $\%$ between $T=2.4$ K, $B=0$ T [Fig.~\ref{fig3}(c)], and $T=25$ mK, $B=15$ T [Fig.~\ref{fig3}(a)]. One scenario might be that there is a $T$-dependent increase of the spin gap due to the exchange interaction, which attempts to prevent electrons entering the spin-up subband. This would cause the compressibility response to be the same for increasing $B$ (which opens the spin gap) and increasing $T$.

In summary, we have performed compressibility measurements on a 1D quantum wire at zero and finite magnetic fields. We resolve spin splitting in the compressibility signal, which, for the lowest subband, is manifest as a peak in $\mathrm{d}V_{sg}/\mathrm{d}V_{mid}$ emerging with increasing field. The temperature response in this regime is remarkably similar, suggesting that the origin of the 0.7 structure is due to spin-polarization. We compare our data to the Kondo scenario, which does not appear to reproduce our results.

This work was supported by the Engineering and Physical Sciences Research Council (UK) and by the Australian Research Council under the DP, LX and APF schemes. K.J.T. acknowledges support from WCU program through the NRF of Korea funded by the Ministry of Education, Science and Technology (R32-2008-000-10204-0). The authors thank O. Klochan for assistance with measurements and thank A.P. Micolich and F. Sfigakis for helpful discussions.


\begin{thebibliography}{MM}
\item $^*$Email address: lws22@cam.ac.uk
\item $^\dag$Email address: kalarikad@skku.edu
\bibitem{thornton86} T. J. Thornton, M. Pepper, H. Ahmed, D. Andrews, and G. J. Davies, Phys. Rev. Lett. \textbf{56}, 1198 (1986).
\bibitem{wharam88} D. A. Wharam \emph{et al.}, J. Phys. C \textbf{21}, L209 (1988).
\bibitem{vanWees88} B. J. van Wees \emph{et al.}, Phys. Rev. Lett. \textbf{60}, 848 (1988).
\bibitem{thomas96} K. J. Thomas \emph{et al.}, Phys. Rev. Lett \textbf{77}, 135 (1996).
\bibitem{wang96} C.-K. Wang and K.-F. Berggren, Phys. Rev. B \textbf{54}, R14257 (1996).
\bibitem{cronenwett02} S. M. Cronenwett \emph{et al.}, Phys. Rev. Lett. \textbf{88}, 226805 (2002).
\bibitem{meir02} Y. Meir, K. Hirose, and N. S. Wingreen, Phys. Rev. Lett. \textbf{89}, 196802 (2002).
\bibitem{klironomos06} A. D. Klironomos, J. S. Meyer, and K. A. Matveev, Europhys. Lett. 74, 679 (2006).
\bibitem{matveev04} K. A. Matveev, Phys. Rev. Lett. \textbf{92}, 106801 (2004).
\bibitem{reilly02} D. J. Reilly \emph{et al.}, Phys. Rev. Lett. \textbf{89}, 246801 (2002).
\bibitem{bird04} J. P. Bird, and Y. Ochiai, Science \textbf{303}, 1621 (2004).
\bibitem{appleyardPRB2000} N. J. Appleyard \emph{et al.}, Phys. Rev. B \textbf{62}, 16275 (2000).
\bibitem{molenkampPRL1990} L. W. Molenkamp, H. van Houten, C. W. J. Beenakker, R. Eppenga, and C. T. Foxon, Phys. Rev. Lett. \textbf{65}, 1052 (1990).
\bibitem{reznikovPRL95} M. Reznikov, M. Heiblum, H. Shtrikman, and D. Mahalu, Phys. Rev. Lett. \textbf{75}, 3340 (1995).
\bibitem{rochePRL04} P. Roche \emph{et al.}, Phys. Rev. Lett. \textbf{93}, 116602 (2004).
\bibitem{eisenstein92} J. P. Eisenstein, L. N. Pfeiffer, and K. W. West, Phys. Rev. Lett. \textbf{68}, 674 (1992).
\bibitem{eisenstein94} J. P. Eisenstein, L. N. Pfeiffer, and K. W. West, Phys. Rev. B \textbf{50}, 1760 (1994).
\bibitem{martinNature08} J. Martin \emph{et al.}, Nat. Phys. \textbf{4}, 144 (2008).
\bibitem{Drexler1994} H. Drexler \emph{et al.}, Phys. Rev. B \textbf{49}, 14074 (1994).
\bibitem{Smith1987} T. P. Smith III \emph{et al.}, Phys. Rev. Lett. \textbf{59}, 2802 (1987).
\bibitem{castleton} I. M. Castleton \emph{et al.}, Physica B \textbf{249-251}, 157 (1998).
\bibitem{luscher} S. L\"{u}scher \emph{et al.}, Phys. Rev. Lett. \textbf{98}, 196805 (2007).
\bibitem{thomas99} K. J. Thomas \emph{et al.}, Phys. Rev. B \textbf{59}, 12252 (1999).
\bibitem{fixedG2} We choose $G_2$ to be in a region of high transconductance (i.e. on a riser between plateaux), so that $\delta E_p$ is amplified by large changes in $G_2$. 
\bibitem{sensitiv} The sensitivity of the measurement is higher for low $G_2$~\cite{castleton}. Therefore data is presented for G2 at $1.5(2e^2/h)$.
\bibitem{corr} To aid comparison, traces in Fig.~\ref{fig3}(c) and (d) were horizontally shifted to correct for a small drift in the pinch-off voltage, since measurements were performed over a 2 week period.
\bibitem{G2} $G_2=0.23(2e^2/h)$ was chosen because below $e^2/h$ the gradient of the conductance does not change significantly with magnetic field or temperature. Therefore the measurement sensitivity remains constant for all data.
\bibitem{noteDFT} In Ref.~\cite{luscher}, local spin density approximations (LSDA) were used to model the compressibility response of a quasibound state in the wire, necessary for the Kondo effect. The screened potential ($V_{det2}$) from biased surface gates ($V_{QPC}$) is detected 200 nm below the 2DEG, and the derivative $D^{DFT}\equiv \mathrm{d}V_{det2}/\mathrm{d}V_{QPC}$ is equivalent to our $\mathrm{d}V_{sg}/\mathrm{d}V_{mid}$ measurement.
\bibitem{highT} We have also compared our measured data at $T=2.4$ K with the DFT ($T = 0$) calculations. At $T=2.4$ K there is a strong 0.7 structure in the conductance and an associated peak in the compressibility signal. However, this compressibility feature is qualitatively different from that predicted as a result of the Kondo effect.

\end{thebibliography}
\end{document}